\documentclass[aps,pre,twocolumn,showpacs,floatfix]{revtex4}
\usepackage{epsfig}
\usepackage{amsmath}
\usepackage{amssymb}
\usepackage{amsfonts}
\usepackage{graphicx}
\usepackage{subfigure}

\newcommand     {\beq}[1]         { \begin{equation} #1 \end{equation} }
\newcommand     {\beqa}[1]        { \begin{eqnarray} #1 \end{eqnarray} }

\newcommand     {\mean}[1]        {\langle #1 \rangle}
\newcommand     {\EP}             { \varepsilon }
\newcommand     {\SI}             { \sigma }
\newcommand     {\EC}             { \varepsilon_c }
\newcommand     {\SC}             { \sigma_c }

\newcommand     {\PPE}             { P(\varepsilon) }

\newcommand     {\KM}             { k_{\textrm{max}} }
\newcommand     {\PPKE}             { P_k(\varepsilon) }

\newcommand{\picfac}{0.995 }

\begin{document}

\title{Continuous Damage Fiber Bundle Model for Strongly Disordered Materials}

\author{Frank Raischel${}^1\footnote{Electronic 
address: raischel@icp.uni-stuttgart.de}$, Ferenc Kun${}^{2}$, 
Hans J.\ Herrmann${}^{1,3}$}  
\affiliation{
\centerline{${}^1$ICP, University of Stuttgart, Pfaffenwaldring 27, D-70569
Stuttgart, Germany}
\centerline{${}^2$Department of Theoretical Physics, University of Debrecen, P.\
O.\ Box:5, H-4010 Debrecen, Hungary}  \\
\centerline{${}^3$Computational Physics, IfB, HIF E12, ETH H\"onggerberg,  CH-8093 Z\"urich, Switzerland}
} 

\date{\today}
             
\begin{abstract}

We present an extension of the continuous damage fiber bundle model to
describe the gradual degradation of highly heterogeneous materials
under an increasing external load. Breaking of a fiber in the model is
preceded by a sequence of partial failure events occurring at random
threshold values. In order to capture the subsequent propagation and
arrest of cracks, furthermore, the disorder of the number of
degradation steps of material constituents, the failure
thresholds of single fibers are sorted into ascending order and their
total number is a Poissonian distributed random variable over the
fibers. Analytical and numerical calculations
showed that the failure process of the system is governed by extreme
value statistics, which has a substantial effect on the macroscopic
constitutive behaviour and on the microscopic bursting activity as
well.

\end{abstract}

\pacs{46.50+a, 62.20.Mk, 81.40.Np}
\maketitle
\section{Introduction}

Materials with a highly disordered microstructure exhibit a variety of constitutive characteristics when subjected to an increasing  external load
---from perfectly brittle to perfectly plastic as well as strain hardening and softening\cite{hh_smfdm,krajcinovic_damage_disorder}. Experiments have
revealed that even if the constituents are brittle, the macroscopic
behaviour shows these variations, especially if the internal structure
of the specimen has a hierarchy of length scales. A possible explanation of this observation is the
gradual activation of failure mechanisms relevant at different length
scales of the materials' microstructure \cite{kun_epjb_2000}. This interaction of microstructure and failure mechanism is typical
for fiber reinforced composites, where fibers of a few micrometer diameter are
embedded in a matrix material \cite{hull_clyne}.  The same geometrical arrangement is
found on a larger scale when pieces of timber are glued together to form
high strength wood structures \cite{timber_eng}. In such materials the microscopic
origin of gradual degradation is the accumulation of damage in fibers and timber pieces due to the
growing population of  microcracks \cite{guarino_microcrack,zapperi_crackfuse_eujb_2000}, furthermore, the growth
and arrest of larger cracks spanning over several length scales of the
specimen \cite{PhysRevB.55.11270,charles:333,roux_ejmech_crackarrest}.

Fiber bundle models (FBM) are one of the most adequate approaches to
understand the fracture  of heterogenous materials \cite{Ferenc_india,andersen_tricrup_prl_1997,statmodfract_alava_2006,toussaint_physa_312_159}. Recently a
continuous damage fiber bundle model (CDFBM) has been introduced where
the stiffness of fibers is reduced in subsequent failure events
representing the gradual degradation mechanism of the material \cite{kun_epjb_2000,raul_burst_contdam,zap_nature_1997}. Recent experiments
revealed that due to the highly heterogeneous microstructure of
faults, defects (like the knotholes of wood, \cite{niemz_holzphysik}), and also particularly
failure resistant spots, the maximum number of failure of composites'
constituents show strong variations inside samples, which  has also an
important effect on the evolution and propagation of cracks. Since the stress concentration around the tip of the crack increases with the crack
length, cracks that grow larger than a certain critical size are unstable \cite{hh_smfdm}. However, cracks can also become arrested at strong locations inside the
material, where the stress concentration around the growing crack demands an increasing strength of subsequent arrest locations \cite{green_crackarrest99,charles:333,roux_ejmech_crackarrest}. This phenomenon therefore requires the knowledge of the distribution of the strongest points in the sample, an issue that is related to
order statistics \cite{hansen_cism_toolbox,randles_nonparstat}. 

In this paper we propose an extension of CDFBM in order to capture the
experimentally relevant failure mechanisms of composite materials
discussed above. We consider  a bundle of parallel fibers which
undergo a sequence of failure events when the local load on them
exceeds the corresponding threshold values. The maximum number of
allowed failures is a Poisson distributed random variable in the model
which accounts for the disorder of degradation steps of fibers
in composites. Since growing cracks sweep over locations of increasing
strength, in the model the breaking thresholds of consecutive
failure events of single fibers are sorted in ascending order. We
explore the complexity of the model by analytical and numerical
calculations and show that the extreme value statistics of failure thresholds has a substantial effect on the fracture process. Contrary to the conventional CDFBM, the macroscopic response of the system does not have a plastic plateau, but instead strain hardening occurs. Computer simulations revealed that on the microlevel the burst size distribution follows a power law with a crossover of the  exponent from $5/2$ to $3/2$ when changing the parameters of the model. Beyond the general interest, our model can in particular be applied to describe the stability of large scale composite wood structures. Following the catastrophic collapse of several buildings composed of such structures in the winter 2006/2007, see e.~g.~\cite{reichenhall01}, this issue has received growing interest in the recent past.

\section{Randomly distributed failure numbers}\label{sec:cont:poiss}

As a model  we employ a continuous damage fiber bundle model (CDFBM), in which the damage law of the classical dry fiber bundle model (DFBM) is supplemented by a gradual degradation of fiber strength in the sequence of failure events \cite{kun_epjb_2000,raul_burst_contdam}. It was shown in  Ref.~\cite{raul_burst_contdam} that for certain choices of the model parameters a variety of experimental situations can be recovered, i.e. either strain hardening or plasticity can occur. On the micro-scale, the size distribution of avalanche events shows a power law behaviour, but the exponent is different from the ordinary FBM \cite{kloster_pre_1997}, and for certain choices of parameters an exponential cutoff appears \cite{raul_burst_contdam}.

  The  CDFBM is constructed as follows: the bundle consists of $N$ parallel fibers  on a square lattice with identical Young-modulus $E$ and random failure thresholds $\sigma_{th}^i$, $i=1, \ldots , N$ with a  probability density $p$ and distribution function $P$. 
Under loading, the fibers  behave linearly elastic until they reach their respective  points of failure and break in a brittle manner, i.e. as soon as the load on a fiber  exceeds 
its breaking threshold $\sigma_{th}^i$, the fiber will fail. 
If the external strain is kept fixed (strain-controlled loading), then the fibers break one by one in  the order of their breaking thresholds, and the full constitutive curve is explored. Under stress controlled loading, however, the load on a breaking fiber is redistributed on the remaining intact ones, where it can trigger subsequent avalanches of fiber breakings \cite{sornette_jpa_1989}. At the maximum of the constitutive curve, a catastrophic avalanche then results in failure of the entire system \cite{hansen_distburst_local_1994,kloster_pre_1997}.
\begin{figure}
  \begin{center}
    \includegraphics[width= .7 \linewidth ,clip, bb= 134 222 393 671]{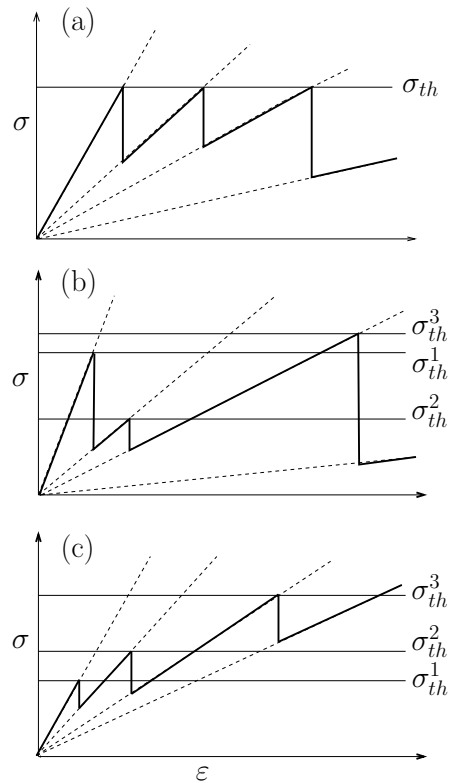}
    \caption{Damage law for a single fiber for the cases of quenched (a), annealed (b) disorder, cf.~\protect{\cite{kun_epjb_2000}}, and for the model with sorted thresholds (c) discussed in Sec.~\ref{sec:cont:sorted}. In all cases, after breaking of a fiber upon reaching a failure threshold $\sigma_{th}^i$, loading is resumed with a stiffness that is reduced by a factor $a$.  }
    \label{fig:cont:mydamlaw}
  \end{center}
\end{figure}
 
 The failure law of the DFBM is  modified in the CDFBM by assuming that at the failure point the Young modulus of the fiber is reduced by a factor $a$, where $0 \leq a < 1$; consequently the stiffness of the fiber after failure is $aE$. The loading of the fiber will then  resume in a linear manner with the reduced stiffness until the next breaking threshold is reached. The parameter $\KM$ determines the maximum number of failures allowed for a single fiber.  The damage threshold $\sigma_{th}^i$ can either be kept constant for all the  breakings (quenched disorder) or new failure thresholds of the same distribution can be chosen (annealed disorder) after each instant of failure, which can model a microscopic rearrangement of the material after failure, cf.~cases (a) and (b) of Fig.~\ref{fig:cont:mydamlaw} \cite{nukala_nacre_pre,turcotte_damrheol_continent,raul_chains_prl,moreno_contdam_time_jphys,moreno_contdam_time_pre}. 

It can be assumed that in an actual experimental situation, the number of times that a constituent of the material can break is an independent realization of an integer  random variable. A prime example is the fracture of wood, specifically of glued timber \cite{timber_eng,hernandez_glulam}, where only a few large defects and the finite number of glued joints determine the extreme value statistics that governs the propagation and arrest of cracks.
This fact can be incorporated  by modeling  $\KM$ as a random number, which is governed by a Poissonian distribution
\beq{ \label{eq:cont:poissonian}
  n_{\kappa} (\KM) = \frac{\kappa^{\KM } e^{-\kappa}}{\KM !} \,.
  }
A new parameter then enters the model, which is the mean value of $\KM$, $\kappa = \mean{ \KM } $.

With this prescription, the macroscopic response of the system can be expressed as
\begin{equation} \label{eq:cont:cb1}
  \SI = \EP \sum_{\KM =0}^{\infty} \frac{\kappa^{\KM } e^{-\kappa}}{\KM !}  \left[ \sum_{k=0}^{\KM } a^k \PPKE \right],
\end{equation}
i.~e.~the Poissonian distribution is convoluted with the degradation term  obtained  for the constitutive behaviour in the continuous damage model \cite{kun_epjb_2000}. In the case of   quenched disorder for the failure thresholds,  the probabilities $\PPKE$ that at a given deformation $\EP$ a randomly chosen fiber has failed exactly $k$ times is
\beq{ \label{eq:cont:quenchedP}
  \PPKE=\begin{cases}
    1-\PPE&, k =0;\\
    P(a^{k-1} \EP) - P(a^{k} \EP)&, 1\leq k \leq \KM ;\\
    P(a^{\KM -1} \EP)&, k=\KM \,.
  \end{cases}
}
For the following discussion, the damage thresholds  will be drawn exclusively from a Weibull distribution with $\lambda=1, m=2$, unless otherwise mentioned.

We can  also apply this model to the case of annealed disorder, where
\beq{ \label{eq:cont:annealedP}
  \PPKE=\begin{cases}
    [1-P(a^k \varepsilon)]\prod_{j=0}^{k-1} P(a^j \varepsilon ) & ,  0 \leq k \leq \KM -1;\\
    \prod_{j=0}^{\KM -1} P(a^j \varepsilon ) & ,  k = \KM \, ,
  \end{cases}
}
however, we restrict this discussion to the  quenched disorder case. In Sec.~\ref{sec:cont:sorted},  a new concept will be introduced which to a certain extent describes a third alternative to  the cases of quenched and annealed disorder.

\begin{figure}
  \begin{center}
    \includegraphics[width=\picfac \linewidth ,clip, bb= 20 27 365 311]{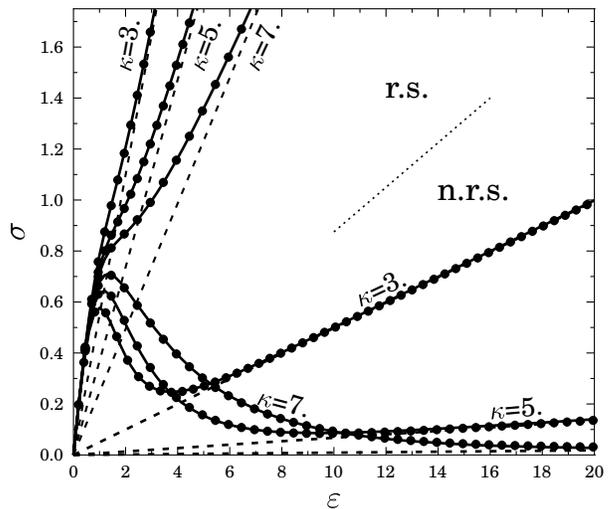}
    \caption{Constitutive behaviour for a fiber bundle with quenched disorder and Poisson distributed $\KM$ for three values of $\kappa$, with (r.s.) and without (n.r.s.) residual stiffness. Symbols: analytical solution, Eq.~(\ref{eq:cont:cb1}), solid lines: simulation results.  The dashed lines show the  asymptotic behaviour in the presence of residual stiffness, Eq.~(\ref{eq:cont:quenched_res_asymp}), and without, Eq.(\ref{eq:cont:quenched_nores_asymp}).}
    \label{fig:var:qconst}
  \end{center}
\end{figure}

In the analytical solution for the constitutive behaviour, Eq.~(\ref{eq:cont:cb1}), two physically strongly distinct cases can be realized by appropriate choices of the summation limit of the innermost, bracketed term: if the summation extends from zero to $\KM$, as indicated, the fibers will retain a residual stiffness after the limiting case of $\KM$ failures, i.e. hardening of the fiber bundle occurs in the limit of large $\EP$, and the asymptotic behaviour of the bundle is described by
\beqa{ \label{eq:cont:quenched_res_asymp}
  \sigma_{\textrm{asympt.}} &=&   \EP \sum_{\KM =0}^{\infty} \frac{\kappa^{\KM } e^{-\kappa}}{\KM !}  a^{\KM}     \nonumber \\
  &=& \varepsilon e^{-\kappa (1-a)} \,.
  }
Fig.~\ref{fig:var:qconst} demonstrates the  perfect agreement between the analytical solution, Eq.~(\ref{eq:cont:cb1}), and a strain controlled numerical simulation, where the asymptotic linear behaviour of Eq.~(\ref{eq:cont:quenched_res_asymp}) is recovered. In order to model the failure of materials, however, the failure law has to be slightly modified: after $k^{\ast}=\KM -1$ failures, the load on a fiber must be set to zero, and the constitutive behaviour changes to
\begin{equation} \label{eq:cont:cb1_nores}
  \SI = \EP \sum_{\KM =0}^{\infty} \frac{\kappa^{\KM } e^{-\kappa}}{\KM !}  \left[ \sum_{k=0}^{k^{\ast} } a^k \PPKE \right] \,.
\end{equation}
Again, in this case the constitutive curves displayed in Fig.~\ref{fig:var:qconst} show an excellent agreement between the analytical solution Eq.~(\ref{eq:cont:cb1_nores}) and the simulation data. At first glance, the hardening behaviour that emerges after the main course of loading may appear to contradict the fact that residual stiffness is not explicitly taken into account and the terms $\PPKE$ with $k=\KM$ are excluded in the failure law, Eq.~(\ref{eq:cont:cb1_nores}); this regime is dominated by the fibers with $\KM=0$, i.~e.~, fibers that never break, and since the expectation value of these fibers is $n_{\kappa} (\KM=0)=e^{-\kappa}$, the asymptotic behaviour even in the case without an explicit residual stiffness reads
\beq{ \label{eq:cont:quenched_nores_asymp}
  \sigma_{\textrm{asympt}} = \varepsilon e^{-\kappa}  \,.
  }
Apparently, the dominance of the fibers with $\KM=0$ diminishes with increasing $\kappa$, and at least for this choice of the disorder distribution, under stress controlled loading the hardening regime cannot be accessed, i.e. all fibers break before traversing the local minima of the slope.  One should note that ---with or without a residual stiffness term--- the fibers with vanishing $\KM$ can be excluded from both the simulations and the analytical calculations, in case of which the hardening behaviour in the second case will also disappear. The distribution $n_{\kappa} (\KM)$ will not be purely Poissonian anymore, and an account of this situation is planned for a future publication \cite{raischel_holz}.

Concerning the avalanche size distribution, this model reproduces the behaviour of  the case of fixed $\KM$, which has been discussed in \cite{raul_burst_contdam}. There, it was found that for larger values of $a$, corresponding to $a > 0.3$ for the Weibull distribution, the distribution can be fitted to a power law, the  exponent of which also depends on $\KM$. For small values of $\KM$, the usual mean field behaviour with an exponent $-5/2$ is obtained \cite{kloster_pre_1997}, whereas for larger values of $\KM$ a smaller exponent $\approx 2.12$ appears. 
\begin{figure}
  \begin{center}
    \includegraphics[width=\picfac \linewidth ,clip, bb= 20 27 365 311]{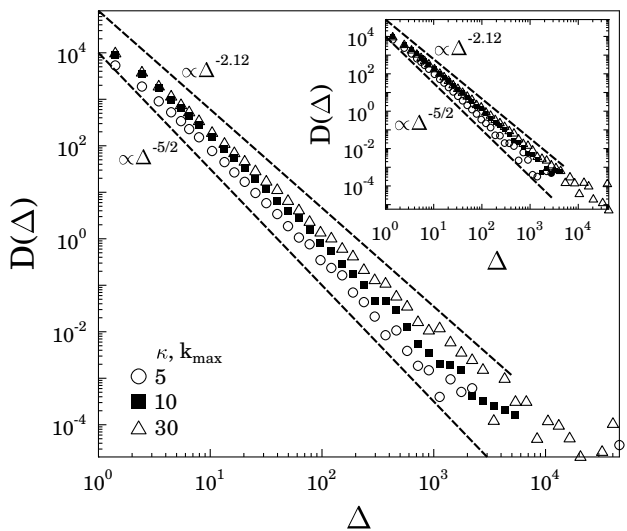}
    \caption{Avalanche size distribution for a fiber bundle of size $L=201$ with  Poisson distributed $\KM$ for different values of $\kappa$. Inset: Avalanche size distribution for a fiber bundle of size $L=201$ with  fixed $\KM$, where the $\KM$ values  shown  correspond to the $\kappa$ values of the main plot.}
    \label{fig:cont:poisavals}
  \end{center}
\end{figure}
Examining the effect of the Poissonian term with a  choice of $\kappa = \mean{ \KM }$ corresponding to the ordinary CDFBM, we find in Fig.~\ref{fig:cont:poisavals} a quantitative agreement between the two models, i.~e.~the Poissonian term causes no visible change to the avalanche statistics, and even the crossover between the two  power law exponents is recovered.

\section{Sorted failure thresholds}\label{sec:cont:sorted}

In highly disordered materials subjected to an increasing external load,  cracks nucleate in the early stages of loading at the weakest locations in a spatially random manner. As the load increases, simultaneously to the nucleation of new microcracks the existing cracks propagate and become unstable. Advancing cracks can be arrested by high strength locations of the material. Before macroscopic failure occurs, advancing cracks can undergo several activation and arrest events. Since stress concentration at the crack tip increases as the crack becomes longer \cite{hh_smfdm}, arresting can only be realized by local material elements of increasing strength. These growth and arrest events result in a gradual degradation of the macroscopic sample strength. 

In order to provide a more realistic representation of this gradual degradation process sweeping through material elements in the increasing order of their local strength, we further extend the CDFBM by  sorting the activation thresholds into increasing order. Fig.~\ref{fig:cont:mydamlaw}, case (c), provides a graphical illustration of the sorting and the ensuing damage law for a single fiber. It is important to emphasize that from  a physical point of view this case is a mixture of the annealed and quenched disorder cases discussed previously \cite{kun_epjb_2000}. On the one hand, the sorted model bears resemblance to annealed disorder, since the consecutive thresholds are different from each other; on the other hand, it could also be classified as quenched as the thresholds are fixed in advance. Sorting of a series of random numbers imposes a correlation between these numbers, and we will have to resort to a mathematical theorem from the field of order statistics in order to obtain a complete understanding of the failure mechanism. It should be stressed that in the following discussion we will consider $\KM$ to assume  a fixed value,  although the addition of the randomly distributed $\KM$ can be trivially incorporated.
Here a single fiber ---seen as a meta-element representing smaller constituents---  models the progress of a crack, which does not proceed continuously but comes to a halt at certain values of the fibers' strain $\varepsilon$.  We may apply the damage law of the CDFBM, but impose the additional condition that the load on the fiber at subsequent instants of the arrest should increase. Hence we can draw the failure thresholds from a random distribution, for which we will use again the Weibull distribution, and store them in sorted order. As in the previous discussion, the Weibull distribution  employed will have the parameters  $\lambda=1$ and $m=2$.

\subsection{Macroscopic Response}\label{sec:cont:sorted:macro}

As mentioned above, bringing an array of $n$ random numbers in sorted order necessarily invokes correlations between them, and the distribution of the random variant at the $i$th position is not governed by the PDF of the unsorted random numbers anymore. 

\begin{figure}
  \begin{center}
    \includegraphics[width=\picfac \linewidth ,clip, bb= 20 27 365 311]{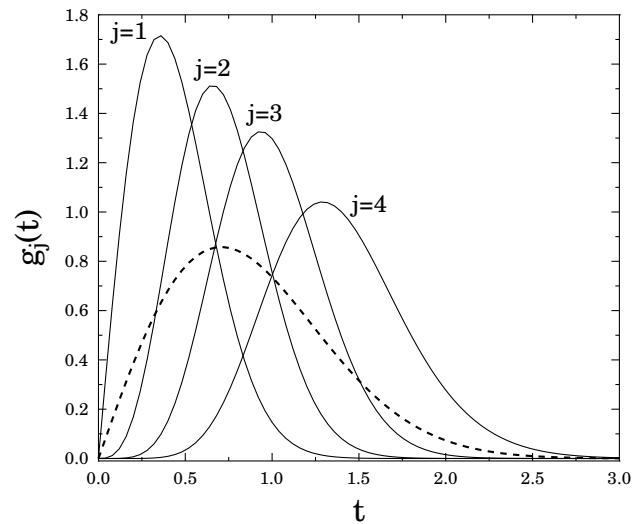}
    \caption{Marginal densities for the order statistics of a Weibull distribution with $\lambda=1, m=2$, for the respective values $j=1, \dots, n=4$ (solid lines), together with the underlying PDF (dashed line), cf.~Eq.~(\protect{\ref{eq:cont:marginalstat}}). }
    \label{fig:cont:marginalstat}
  \end{center}
\end{figure}
From order statistics, the following results pertaining to a sequence of random numbers $X_1,\dots ,X_n$ drawn from a CDF $F(x)$ and density $f(x)$ are known \cite{randles_nonparstat}: if the $X_i$ are brought into increasing order, then the numbers $X_{(i)} \leq \cdots \leq X_{(n)}$ are called the order statistics of the $X_i$, which have the joint density
\beq{
    g(x_{(1)},\dots,x_{(n)}) = 
  \begin{cases}
    n! \prod \limits_{i=1}^{n} f(x_{(i)}), \\
    \qquad  -\infty < x_{(1)} < \dots < x_{(n)}< \infty \\
    0, \\
    \qquad \mathrm{elsewhere} \, .
    \end{cases}
  }
The marginal density for the $j$th order statistics $X_{(j)},$  $1 \leq j \leq n$ is
\beqa{\label{eq:cont:marginalstat}
  g_{(j)} (t) &=&   \frac{n!}{(j-1)! (n-j)!} [ F(t)]^{j-1} [1- F(t)]^{n-j} f(t), \nonumber \\
  && -\infty < t < \infty \, .
  }

The marginal statistics Eq.~(\ref{eq:cont:marginalstat}) is therefore the adequate replacement of the PDF, i.~e.~the distribution function for the random number at the $j$-th position, if  $n$ random numbers have been drawn. In order to illustrate this result,  in Fig.~\ref{fig:cont:marginalstat} the marginal statistics for the $j$th random number, $1 \leq j \leq n=4$, together with the underlying Weibull distribution with $\lambda=1, m=2$ is shown. 
It is apparent from Fig.~\ref{fig:cont:marginalstat} that with increasing $j$, the marginal statistics share the peaked characteristics of the underlying PDF, and that the position of the maxima reflects the sorting. Also, the marginal statistics become wider with increasing $j$, although this effect is not too pronounced.
\begin{figure}
  \begin{center}
    \includegraphics[width=\picfac \linewidth ,clip, bb= 20 27 365 311]{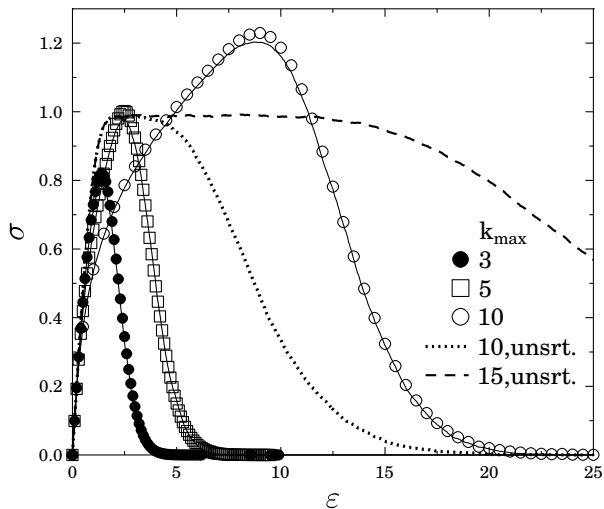}
    \caption{Constitutive curves for the model with  sorted random thresholds, without residual stiffness. Solid lines: simulation results, symbols: analytical solution Eq.~(\ref{eq:cont:cc_as}), dashed and dotted lines: simulations for the conventional CDFBM without sorting, shown for comparison. The failure thresholds are drawn from a Weibull distribution with $\lambda=1,m=2$.}
    \label{fig:cont:cc_as}
  \end{center}
\end{figure}

Having found an analytical expression for the marginal statistics,  the constitutive behaviour of this model can be expressed in a closed form. In analogy to the annealed case of the CDFBM, we denote by $P_k (\EP)$ the probability that a fiber has failed exactly $k$ times at a strain $\varepsilon$:
\beq{\label{eq:cont:pkas}
  P_k (\EP) = \begin{cases}
    [1-G_{(k+1)}(a^k \EP) ] \prod \limits_{j=0}^{k-1} G_{(j+1)}(a^j \EP),\\
    \qquad 0 \leq k \leq \KM -1\\
    \prod \limits_{j=0}^{\KM -1} G_{(j+1)}(a^j \EP),\\
    \qquad k = \KM \, ,
    \end{cases}
    }
    
\noindent where 
\beq{\label{eq:cont:Gj}
  G_{(j)} (x) = \int_{0}^{x} g_j(t) \, dt
}
is the integral associated with the marginal statistics $g_{(j)} (t)$, corresponding to the CDF of unordered random numbers.
The second case in Eq.~(\ref{eq:cont:pkas}) with $k = \KM$ corresponds to the residual stiffness of the bundle. With this result the analytic solution for the constitutive behaviour reads 
\beq{ \label{eq:cont:cc_as}
  \sigma(\EP ) = \sum \limits_{k=0}^{\KM -1} a^k \EP [1-G_{(k+1)}(a^k \EP) ] \prod \limits_{j=0}^{k-1} G_{(j+1)}(a^j \EP) \, ,
  }
if the hardening term is skipped in order to account for material failure.

It should be noted, though, that in the formula for the $k$-th failure probability $P_k (\EP)$, Eq.~(\ref{eq:cont:pkas}), the integral Eq.~(\ref{eq:cont:Gj}) appears, which cannot in general be solved analytically due to the structure of the integrand, Eq.~(\ref{eq:cont:marginalstat}). In order to  obtain the constitutive behaviour   the integral Eq.~(\ref{eq:cont:Gj}) has to be evaluated numerically. In Fig.~\ref{fig:cont:cc_as}, the stress-strain curves obtained in this way are plotted for  three values of $\KM$, where an excellent agreement between the numerical and analytical results, Eq.~(\ref{eq:cont:cc_as}) can be observed. 
\begin{figure}
  \begin{center}
    \includegraphics[width=\picfac \linewidth ,clip, bb= 25 420 391 707]{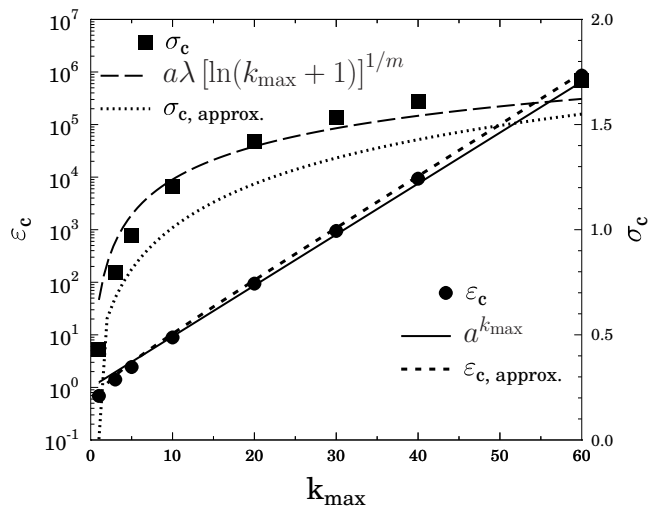}
    \caption{The critical stress $\sigma_c$ and strain $\varepsilon_c$ for a bundle with sorted random thresholds from  a Weibull distribution with $\lambda=1,m=2$  and without residual stiffness as a function of  the maximum number $\KM$ of failures. The symbols represent simulation results, whereas the lines display evaluations of Eqs.~(\ref{eq:cont:sc_approx},\ref{eq:cont:ec_approx}) and numerically obtained  maxima of Eq.~(\ref{eq:cont:ripple_maxima}) with the setting $k=\KM$, respectively.}
    \label{fig:cont:sec}
  \end{center}
\end{figure}
\begin{figure}
  \begin{center}
    \includegraphics[width=\picfac \linewidth ,clip, bb= 37 51 412 334 ]{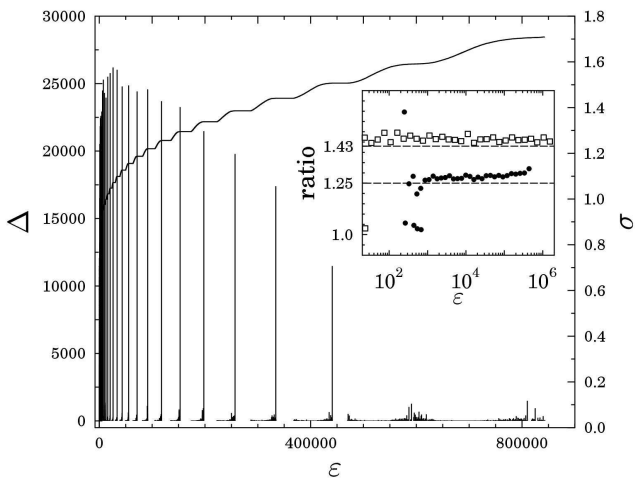}
    \caption{Constitutive curve (solid line) and avalanche sizes without residual stiffness as a function of strain $\varepsilon$  from a simulation with $\KM=60$, $m=2$. Inset: ratio $\EP_{i+1} / \EP_i$ for subsequent bursts with $\Delta_i >200$, for the values $a=0.7$ (unfilled circles) and $a=0.8$ (filled squares), where the lines indicate the respective values of $1/a$.  }
    \label{fig:cont:as_hist_avals}
  \end{center}
\end{figure}
It can be seen in Fig.~\ref{fig:cont:cc_as} that contrary to the case of unsorted thresholds \cite{raul_burst_contdam} ---whether quenched or annealed disorder is of no importance--- that the constitutive curve does not develop a plateau, it always increases monotonically and has a quadratic maximum where macroscopic  failure occurs. As a consequence of extreme value statistics, with growing $\KM$ the critical stress $\SC$ and strain $\EC$ increase, indicating a higher macroscopic load bearing capacity. It can be seen from the general expression of the constitutive curve, Eq.~(\ref{eq:cont:cc_as}), that the macroscopic failure of the system is  mainly controlled by the largest thresholds whose distribution can be obtained from Eq.~(\ref{eq:cont:marginalstat}), setting $j=\KM$. Analyzing the constitutive behaviour of the system considering only the largest thresholds, $j=\KM$, yields
\beq{\label{eq:cont:sc_approx}
  \SC \approx a \lambda \left[ \ln(\KM +1)\right] ^{1/m} 
  }
and 
\beq{\label{eq:cont:ec_approx}
  \EC \approx a^{-\KM} \, ,
  }
for the failure stress and strain, respectively, assuming Weibull distributed failure thresholds with parameters $\lambda, m$. Eq.~(\ref{eq:cont:sc_approx}) implies that the sorted CDFBM does not have a plastic limit as  in the conventional CDFBM, i.~e.~no plateau of the $\sigma (\EP)$ emerges. Instead, the strength of the bundle is an asymptotically increasing function of $\KM$, namely, $\SC$ increases logarithmically whereas $\EC$ increases exponentially with $\KM$. This is illustrated in Fig.~\ref{fig:cont:sec}, where the values $\SC$ and $\EC$ obtained by computer simulations are compared to the analytical results, Eqs.~(\ref{eq:cont:sc_approx},\ref{eq:cont:ec_approx}).

For very high values of $\KM$, a distinguished regime of oscillations appears in the constitutive curves, see Fig.~\ref{fig:cont:as_hist_avals}. This plot shows the constitutive behaviour in a stress controlled simulation with $\KM=60$, together with the avalanche sizes $\Delta$ recorded at each loading step. Apparently, the constitutive curve displays a large amount of oscillations with horizontal plateaus, which coincide with large scale bursts of breaking events. The position of the peaks suggests a regularity of some kind. In order to quantify this regularity of the peak events, the inset carries information about the ratios $\varepsilon_{i+1}/\varepsilon_i$ of subsequent failure events $\Delta$ with $\Delta_i >2000$. Obviously, this ratio assumes a constant value of $\approx 1/a$ after a brief onset period, where $a=0.8$ is the load reduction parameter used in all the simulations presented in this discussion, and a comparison with the case of $a=0.7$ is presented in order to confirm the influence of the load reduction parameter. It should also be noted that  the envelope of the constitutive curve remains monotonically increasing.  In order to analyze this oscillation phenomenon  an investigation on the evolution of the breakdown process is necessary.

\begin{figure}
  \begin{center}
    \includegraphics[width=\picfac \linewidth ,clip, bb= 35 24 371 313]{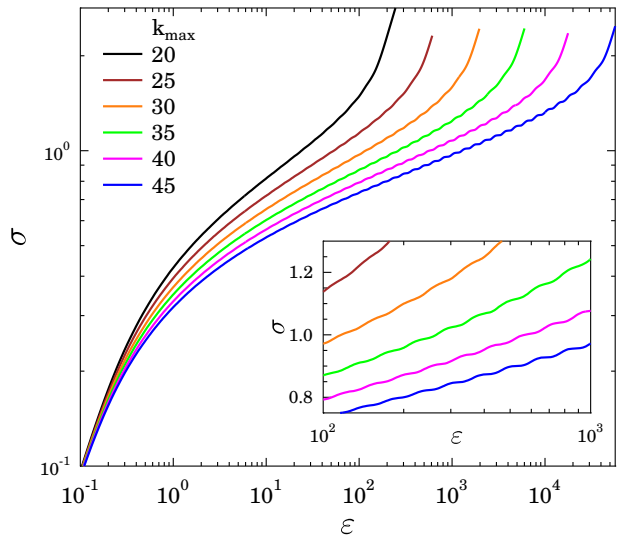}
    \caption{(Color online) Constitutive curve for the model with sorted failure thresholds and residual stiffness with $\lambda=1$, $m=2$ for various values of $\KM$ in a double-logarithmic plot, under stress controlled loading. The curves display the same asymptotic behaviour as the residual stiffness curves in Fig.~\protect{\ref{fig:var:qconst}}, although it is not visible in this logarithmic representation. The inset shows a section of the constitutive curves on a  logarithmic-linear scale, where the onset of oscillations with increasing $\KM$ is better visible. }
    \label{fig:cont:const_oscillations_res}
  \end{center}
\end{figure}
In fact, an analytical argument can be made about the origin of these oscillations, and their position relative to the state of loading $\varepsilon$. In the constitutive formula, Eq.~(\ref{eq:cont:cc_as}),  there appears a product of integral marginal distributions $G_{(j+1)}(a^j \varepsilon)$, where $0 \leq j \leq k-1$, and $0 \leq k \leq \KM -1$. From the structure of Eq.~(\ref{eq:cont:marginalstat}), and if a  Weibull distribution with $m=2, \lambda=1$ is assumed as usual, it can be understood that the respective derivatives $g_{(j+1)} (x)$ take on their  maximum value even for very large $j,n$, and $j \lesssim n$ at a numerical value of the argument  $x = \mathcal{O}(1)$. Therefore, the $g_{(j+1)}$ for large values of $j$ possess well defined peaks  at very large values of $\varepsilon$, where $a^j \varepsilon \approx 1$. The positions of these peaks become strongly separated for subsequent indices $j,j+1$ if $j = \mathcal{O}(n)$ and $n$ is large. Consequently, in Eq.~(\ref{eq:cont:cc_as}) the product of the integral quantities $G_{(j+1)} (a^j \EP)$ can be replaced by the largest factor $G_{(k)} (a^{k-1} \EP)$ for relatively large $k$, and together with the leading term $\varepsilon [1-G_{(k+1)}(a^k \EP) ]  $ a peak structure is formed. It was confirmed numerically for $\KM=60$ and $k \gtrsim 30$ that the positions   $\varepsilon$ of these maxima, which are defined through the function
\beq{\label{eq:cont:ripple_maxima}
 m_k (\EP) =  a^k \EP [1-G_{(k+1)}(a^k \EP) ]  G_{(k)}(a^{k-1} \EP) \, 
}
practically coincide with the observed peaks of avalanches, furthermore,  setting $k=\KM$  also yields good estimates for the critical stress and strain, see Fig.~\ref{fig:cont:sec}.

An oscillatory regime  also appears if residual stiffness is present, as shown in Fig.~\ref{fig:cont:const_oscillations_res} for  stress controlled simulations for various values of $\KM$, where oscillations are clearly visible for $\KM \gtrsim 30$, as in the case without residual stiffness.  Again, the appearance of oscillations depends on the choice of $\KM$, and sets in at $\KM \approx 30$. With increasing $\KM$ also pronounced strain hardening occurs, whereas in the conventional CDFBM a plastic plateau is present. After passing the strain hardening regime, the bundles attain an asymptotic regime of constant slope  $a^{\KM}$ for all values of $\KM$, where also macroscopic failure occurs for all cases investigated. It should be noted that therefore at low values of $\KM$, i.~e.~without oscillations, the slope of the constitutive curve is always finite positive, whereas for higher values of $\KM$, there are sections of the constitutive curve with zero slope due to the oscillations.

\subsection{Bursts of Fiber Breakings}\label{sec:cont:sorted:micro}
\begin{figure}
  \begin{center}
    \includegraphics[width=\picfac \linewidth ,clip, bb= 20 20 365 311]{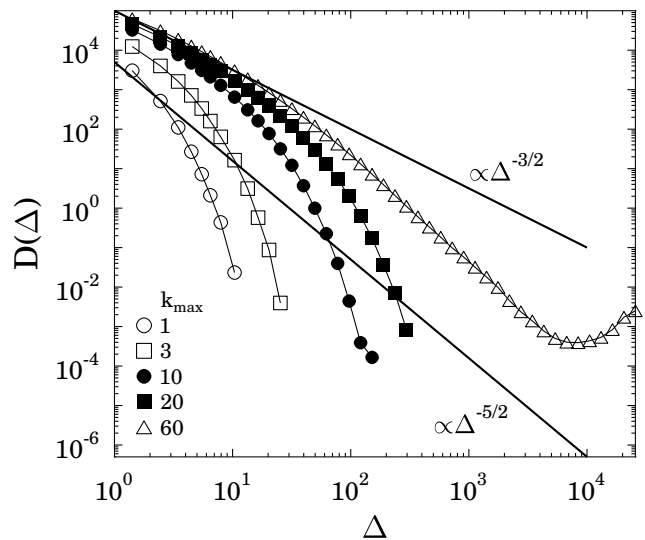}
    \caption{Avalanche size distributions for the CDFBM with sorted failure thresholds for various values of $\KM$, $1 \leq \KM \leq 60$ and residual stiffness. Results from stress controlled simulations of $L=201$ fiber bundles, averaged over  100 realizations.}
    \label{fig:cont:av_as_raw_res}
  \end{center}
\end{figure}
\begin{figure}
  \begin{center}
    \includegraphics[width=\picfac \linewidth ,clip, bb= 20 20 365 311]{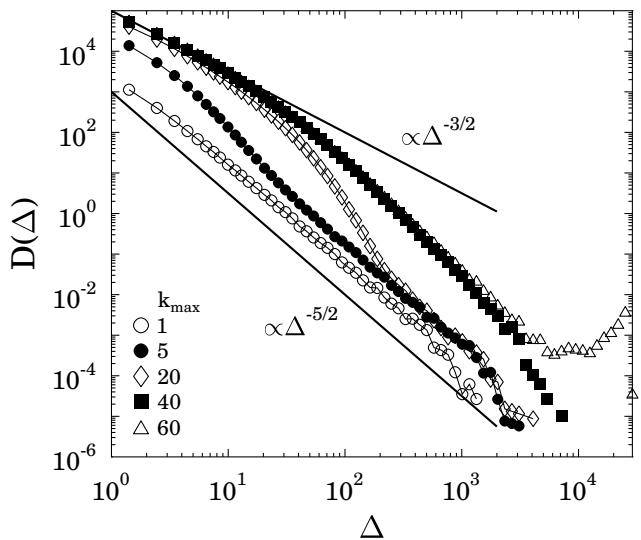}
    \caption{Avalanche size distributions for the CDFBM with sorted failure thresholds and no residual stiffness for various values of $\KM$, $1 \leq \KM \leq 60$. Results from stress controlled simulations of $L=201$ fiber bundles, averaged over  200 realizations.}
    \label{fig:cont:av_as_raw}
  \end{center}
\end{figure}
\begin{figure}
  \begin{center}
    \includegraphics[width=\picfac \linewidth ,clip, bb= 18 411 359 707]{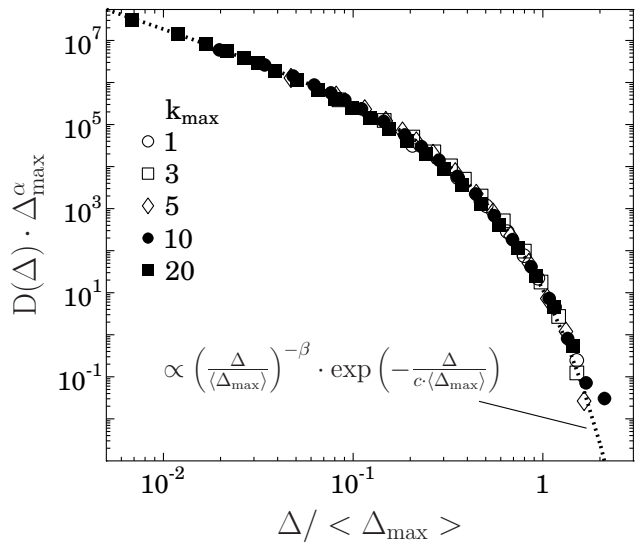}
    \caption{Rescaled avalanche size distributions for the CDFBM with sorted failure thresholds and residual stiffness, for small values of $\KM$ outside the oscillatory regime. A best fit with the numerical values $\alpha=1.22$ and $\beta=1.5$ for the exponents has been used to obtain the collapse of the curves and the dashed fit curve.  }
    \label{fig:cont:av_as_raw_res_scale}
  \end{center}
\end{figure}

The presence of the oscillations, and therefore of a locally vanishing slope, has a distinguished effect on the distribution of avalanche sizes, as demonstrated by Fig.~\ref{fig:cont:av_as_raw_res} for the case with residual stiffness, and in Fig.~\ref{fig:cont:av_as_raw} for the case without; this is also suggested by the presence of large size avalanches in Fig.~\ref{fig:cont:as_hist_avals} (no residual stiffness). It can be seen that in the presence of residual stiffness, Fig.~\ref{fig:cont:av_as_raw_res}, two remarkable features are present. First, for all values of $\KM$, there is a regime with a power law of exponent $-3/2$ for small avalanche sizes. Secondly, for small values of $\KM$, an exponential cutoff appears for the larger avalanches; the presence of the initial power law regime with a following cutoff is confirmed by  rescaling both axes by the average size of the largest avalanche, $\left< \Delta_{\mathrm{max}} \right>$. It is important to note that the  distributions $D(\Delta)$ can  be collapsed onto a single master curve, which was then fit by the functional form
\beq{\label{eq:cont:fit}
  D(\Delta) \propto \left( \frac{\Delta}{\left< \Delta_{\mathrm{max}} \right>} \right)^{-3/2} \exp \left( -\frac{\Delta}{c \left< \Delta_{\mathrm{max}} \right>} \right) \, ,
}
as can be seen in Fig.~\ref{fig:cont:av_as_raw_res_scale}, where the curves with low values $\KM \leq 20$ have been used. The above arguments are supported by the fact  that the power law of exponent $3/2$ with an exponential cutoff provides a perfect fit to the master curve obtained numerically. It has to be emphasized that in the fitting, the value of the exponent was fixed to $3/2$, and the fit was obtained solely by varying the parameter $c$ in Eq.~(\ref{eq:cont:fit}).

For high values of $\KM$, i.~e.~in the presence of oscillations, a crossover is observed from an initial regime with a power law of exponent $-3/2$, to another regime with the mean field power law of exponent $-5/2$, and finally to a peaked regime for very large avalanches of about the system size.

A similarly complex behaviour is observed in the cases without residual stiffness, see Fig.~\ref{fig:cont:av_as_raw}. There, for low values of $\KM$, the usual mean field behaviour of  a power law with an exponent $-5/2$ is observed \cite{hansen_distburst_local_1994,kloster_pre_1997}. For intermediate values of $\KM$, a crossover occurs between an initial $-3/2$ power law part to a mean field part for larger avalanches, and the position of the crossover shifts to larger avalanche sizes with increasing $\KM$. For higher values of $\KM$, a limiting curve with a crossover at $\Delta \approx 10^2$ can be identified, and again a peak of avalanches of the order of the system size  is found for $\KM = 60$. 

\begin{figure}
  \begin{center}
    \includegraphics[width=\picfac \linewidth ,clip, bb= 20 20 365 311]{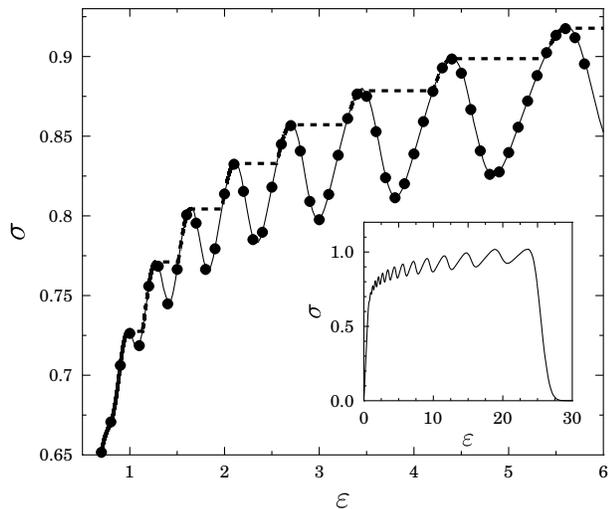}
    \caption{Section of the constitutive curve for the model with $\KM=15$, $m=10$ and no residual stiffness. Solid line: strain controlled simulation, dashed line: stress controlled simulation, dots: exact evaluation of the analytical solution, Eq.~(\ref{eq:cont:cc_as}).}
    \label{fig:cont:const_oscillations}
  \end{center}
\end{figure}

The features found in the avalanche size distributions can all be explained on the basis of the fine structure of the oscillations, which is displayed in Fig.~\ref{fig:cont:const_oscillations} for both stress and strain controlled loading, with $\KM=15$ and the choice $m=10$ for the Weibull parameter. From this illustration it becomes apparent that the horizontal plateaus in the stress controlled simulations actually correspond to regions of decreasing stress $\sigma$ under strain controlled loading, which cannot be accessed in the stress controlled mode. Also, for these parameter values, the integral Eq.~(\ref{eq:cont:Gj}) could be solved analytically by using a computer algebra program, and the analytic solution of Eq.~(\ref{eq:cont:cc_as}) shown in Fig.~\ref{fig:cont:const_oscillations} is in excellent agreement with the simulation results.

In general, a fiber bundle model can only produce large avalanches if the constitutive curve has at least one maximum, where the susceptibility to a small increment of the external force diverges \cite{sornette_jpa_1989,andersen_tricrup_prl_1997}. Avalanches with a power law distribution are generated in the vicinity of the maximum of $\sigma ( \varepsilon) $, where the shape of the maximum determines the value of the exponent $\tau$. Quadratic maxima typically result in $\tau=5/2$, the value obtained in the absence of both oscillations and a residual stiffness term. If loading is stopped at a strain $\EP _s$ before reaching the maximum, i.~e.~before global failure occurs, an exponential cutoff in the avalanche size distribution appears \cite{statmodfract_alava_2006}, which is visible in Figs.~\ref{fig:cont:av_as_raw_res} and \ref{fig:cont:av_as_raw_res_scale}, where due to the residual stiffness term the bundle fails macroscopically after passing exclusively through regions of finite positive slope, without quadratic maxima, as for the cases $\KM \leq 30 $ no oscillations occur. 

 In the oscillatory regime, however, $\sigma ( \varepsilon) $ passes a series of consecutive maxima with an increasing amplitude. Under stress controlled loading the system jumps from a local maximum of $\sigma ( \varepsilon) $ to the ascending side of the next maximum which  is somewhat higher than the previous one, see Fig.~\ref{fig:cont:const_oscillations}. The jump implies that a large  amount of fiber breakings occur  in a single avalanche removing all fibers which have breaking thresholds lower than the load of the ending point of the jump on the next peak of  $\sigma ( \varepsilon) $. Consequently,  when loading is continued along the ascending side of the peak determined by Eq.~(\ref{eq:cont:ripple_maxima}), the response of the system is determined by a disorder distribution which is critical in the sense of Refs.~\cite{hansen_crossover_prl,pradhan_india}, i.~e.~weak fibers are removed so that the lower cutoff of the disorder distribution falls close to the local critical deformation, the location of the next peak. As it has been shown in \cite{hansen_lower_cutoff_2006,raischel_cutoff}, when the disorder distribution approaches criticality, the avalanche size distribution exhibits a crossover from a power law with an exponent $\tau =3/2$ for the small avalanches, to another exponent of $\tau=5/2$ for the large ones, irrespective of the effective range of interaction. This effect can be recognized in the $\KM=60$ curve in Fig.~\ref{fig:cont:av_as_raw_res}.

The same argumentation holds for the case without residual stiffness: here, a global quadratic maximum is always present, so for low values of $\KM$ the mean field exponent $-5/2$ is found. However, in the presence of oscillatory structures, there appear a series of local quadratic maxima, and the stepping effect described above yields a series of local critical threshold distributions, which results in the crossover of exponents visible in the avalanche size distribution, cf.~Fig.~\ref{fig:cont:av_as_raw}.

It has to be stressed that the effects found here in the CDFBM with sorting have no equivalent counterparts in the conventional CDFBM without sorting. There, the macroscopic behaviour yields a plastic plateau and no steps appear in the constitutive curve; consequently, the size distribution of avalanches shows no signatures of criticality \cite{raul_burst_contdam}. The existence of oscillations, synchronized avalanche bursts and a critical crossover of the avalanche size distribution exponents is a genuine peculiarity of the order statistics  accompanying sorting.

\begin{figure}
  \begin{center}
    \includegraphics[width=\picfac \linewidth ,clip, bb= 20 20 365 311]{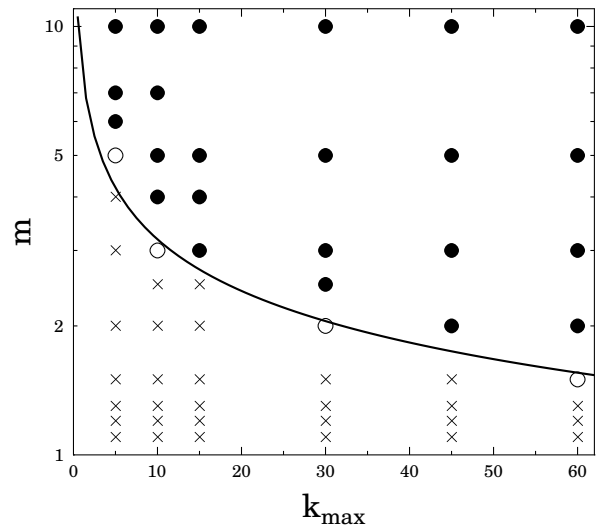}
    \caption{Phase diagram denoting the appearance of oscillations in the constitutive curves of the model with sorting and no residual stiffness, and the presence of pronounced spikes in the avalanche size distribution.  Filled  circles: oscillatory regime; crosses: normal regime; unfilled circles: regimes cannot be distinguished. The solid line denotes the approximate location of the separatrix between the two regimes.}
    \label{fig:cont:phasediag}
  \end{center}
\end{figure}
The analytical argumentation presented above also makes it clear that oscillations cannot appear for all disorder distributions and all values of $\KM$; rather, its appearance is  restricted to combinations of either large $\KM$ and high disorder corresponding to  low values of the Weibull parameter $m$, or smaller values of $\KM$ and correspondingly low disorder, such that the maxima defined through Eq.~(\ref{eq:cont:ripple_maxima}) are clearly separated. Fig.~\ref{fig:cont:phasediag} presents a numerical survey of the apparent occurrence of oscillations. One can see that a well defined and  smooth separatrix can be found which isolates the regime with oscillations from the regime without in the $\{\KM, m\}$ parameter space.  
\section{Conclusions}

Motivated by experimental observations on the fracture process of composite systems having a hierarchy of length scales, we extended the continuous damage fiber bundle model by taking into account that a hierarchy of length and energy scales for damage initiation and crack growth and propagation exists. 
Since macroscopic damage and arrest event are based on large defects and high strength zones in the material, their number is not a constant but varies between samples of the same production batch. Therefore, two new features have been added to the classical CDFBM, and their effect on the microscopic and the macroscopic damage evolution has been investigated. 

First,  the maximum number of failures $\KM$ has been modeled as a Poissonian random variable, which incorporates the existence of disorder 
with respect to the finite number of macroscopic defects and strength zones which  can effectively govern the macroscopic breaking of certain materials. 
The presence of the Poissonian term has  a distinct effect on the constitutive behaviour as it induces a hardening regime, which becomes more pronounced for small average values of $\KM$. For the microscopic behaviour, the introduction of the  Poissonian distributed  $\KM$ leaves the distribution of avalanche sizes invariant,and a crossover from a power law with an exponent $-5/2$ to a power law with another exponent $-2.12$ for increasing values of $<\KM >$ is found, in analogy to the conventional CDFBM. 

In a next step we introduced sorted failure thresholds in the model and explored the inherently complex behaviour in this case by analytical and numerical means.  A parameter regime has been identified where the damage evolution of all fibers synchronizes and considerable changes to the microscopic quantities can be observed, depending on the amount of disorder and the maximum number of allowed failures. It was shown that the extreme value statistics of failure thresholds has a substantial effect on the fracture process of the system both on the micro- and  macro-level. This theoretical study can be particularly important for composite materials produced by assembling components with a large variation in their respective defects. The application of this model to the failure process of large wooden structures is in progress.

\begin{acknowledgments}
We are grateful to M. Schrank for experimental contributions and discussions. This work has been supported by the project SFB 381. F.~Kun acknowledges financial support of the research contracts  OTKA  T049209 and NKFP-3A/043/04.
\end{acknowledgments}

\bibliography{../nachphd}
\end{document}